# A Large-area GEM Detector Using an improved Self-stretch Technique *

YOU Wen-hao(尤文豪)[1]　　ZHOU Yi(周意)[1;1)]　　LI Cheng(李澄)[1]　　CHEN Hong-fang(陈宏芳)[1]
SHAO Ming(邵明)[1]　　SUN Yong-jie(孙勇杰)[1]　　TANG Ze-bo(唐泽波)[1]　　LIU Jian-bei(刘建北)[1]
State Key Laboratory of Particle Detection and Electronics, University of Science and Technology of China, Hefei, 230026, China

**Abstract:** A GEM detector with an effective area of 30 $\times$ 30 $cm^2$ has been constructed using an improved self-stretch technique, which enables an easy and fast GEM assembling. The design and assembling of the detector is described. Results from tests of the detector with 8-keV X-rays on effective gain and energy resolution are presented.

**Key words:** GEM, self-stretch, effective gas gain, energy resolution, uniformity

**PACS:** 29.40.-n

## 1 Introduction

Gas Electron Multiplier (GEM) detectors are widely used as a type of position sensitive gaseous detector for ionizing radiation detection. The work based on a novel concept of charge amplification in gas which was first introduced and then developed by F. Sauli in 1997 at CERN [1]. The charge amplification of a GEM detector arises from GEM foils placed in the detector, they are thin copper-clad kapton foils perforated with a large number of holes by chemical technology. For regular GEM foils, the holes have a double-conical shape With an inner/outer diameter of $50\mu$ m/$70\mu$ m, and are arranged in a hexagon pattern with a hole pitch of $140\mu$ m. When a voltage difference around 400 V is applied on both side of the GEM foil, an electric field as high as 100 kV/cm will be created in the holes and result in the charge amplification [2]. Generally, a GEM detector uses the $Ar/CO_2$ mixture as its working gas, it has a sandwich structure consisting of a drift cathode, two or three GEM foils and a readout anode, as shown in Fig.1.

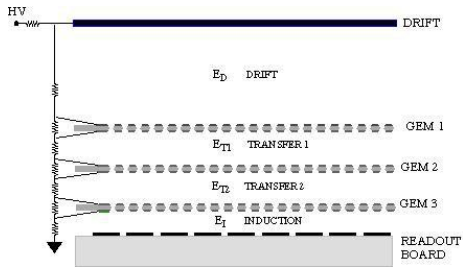

Fig. 1. Schematics of a regular triple-GEM detector

Soon after the advent of the GEM detectors, they found increasing applications in various fields due to their excellent performance with low cost. Among the characteristics of GEM detectors are radiation hardness, high rate capability, good position resolution, flexible detector shape and readout patterns, operation with non-explosive gas mixtures. For a long period of time，the size of GEM foils was limited by the double-mask etching technique(40 $cm$ $\times$ 40 $cm$ maximum) [3], the most frequently used GEM foils are 10 $cm$ $\times$ 10 $cm$. Small GEM detectors using these GEM foils were mainly used in laboratories for position measurement of the charged particles [4,5]. The size limitation of GEM foils had long been a major obstacle preventing GEM detectors from finding applications in large experiments in nuclear and particle physics.

In 2010, a new etching method called single-mask technique was developed at CERN [6]. It overcomes the size limitation arising from the single mask technique, hence greatly increasing the maximum size (2 $m\times$0.6 $m$ [3]) of the GEM foils that can be produced. This makes possible fabrication of large-area GEM detectors which have been intensively used in the areas of nuclear and high-energy physics, medical imaging, radiation monitoring. For example, a set of triple-GEM detectors with an active area of 40 $cm$ $\times$ 50 $cm$ each are used as the tracking device of charged particles for the SBS experiment in Hall-A of Jefferson Lab [7], large-area cylinder-GEM (CGEM) detectors are used as the inner tracker for the KLOE experiment at the DAΦNEΦ-factory[8], and triple-GEM chambers with dimensions of 99 cm$\times$(22 "" 45.5) cm are proposed as a detector option for muon tracking and triggering at the forward region in CMS experiment[9].

4 years ago, Jefferson Lab started their 12 GeV upgrade program in Hall A, it was proposed that large-area

∗ Supported by National Natural Science Foundation of China (11205151, 11375180)
1) E-mail: zhouyi@mail.ustc.edu.cn

GEM detectors with the effective area 1 $m \times 0.4$ $m$ will be used for the inner-tracking system. University of Science and Technology of China (USTC) is a member of the SoLID collaboration and has been involved in the R&D of large-area GEM detectors for the SoLID experiment. As the first step, we has built a 30 $cm \times 30$ $cm$ GEM detector prototype successfully, this is the first R&D work on the large-area GEM detector in China.

## 2 Techniques for assembling large-area GEM detectors

The most critical element in large-area GEM detectors is techniques for stretching GEM foils. For small-area GEM detectors, GEM foils can be directly glued(with very small stretch by hand) on a small frame which can then be placed between the drift cathode and the readout anode. For the large ones, the foils cannot be simply glued on the frame, because the foils would be severely distorted by the electro-static force between electrodes (GEM foils, cathode and anode) and gravity of foils. To construct large-area GEM detectors, several foil-stretching methods have been developed. The first one is a gluing technique developed jointly by CERN and INFN. In this method, a stretching device with tension sensors stretches and holds a large GEM foil, a frame is then glued onto the stretched foil[10]. The second one is a thermal-stretching technique developed by Florida Institute of Technology (FIT). This technique employs Plexiglas frames with a different thermal expansion coefficient than GEM foils. Large-area GEM foils are fixed on the frames and are stretched when the environmental temperature increases resulting in different expansions on the GEM foils and the frames[11].

Some GEM detector prototypes were assembled using the two techniques described above and functioned properly. However, there are some drawbacks of these techniques: 1) the assembling process takes a long time and no mistake is allowed in the process. 2) a GEM detector is integrated into a whole undetachable chamber leaving no possibility of repairing the chamber by replacing any parts. This would unavoidably increase costs and operation risk of GEM detectors.

In 2011, CERN developed a new self-stretch assembling technology called "No Stretch, No Stress (NS2)" technique [12], which is illustrated in Fig.2. In this technique, GEM foils are mechanically fixed by a set of small inner frames, located outside the active area as the first step of the assembly process. There are screws passing though the main frame fastened by nuts embedded in the inner frames, all the foils are tightened. The main frame provides the mechanical tension needed for foil stretching and gas tightness for the detector. Compared with the gluing and thermal stretching techniques, the assembling process with the NS2 technique is much faster and easier. Neither use of glue nor spacer are involved avoids dead areas and improves the gas flow uniformity inside the detector. All the parts of the detector are fastened by screws. So that anything of the detector is replaceable and this greatly decreases the cost of repairmen of the detector.

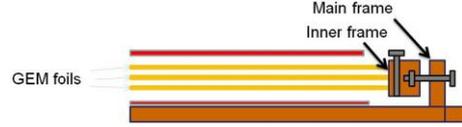

Fig. 2. NS2 self-stretching technique. GEM foils are mechanically fixed by a set of small inner frames. Stretch screws passing though the main frame and were fastened by nuts embedded in the inner frames. The main frame provides the mechanical tension and gas tightness.

## 3 Design and assembling of a $30cm \times 30$ $cm$ GEM detector

Fig.3 shows the design of the prototype, it is a triple GEM detector with a "3-2-2-2" structure, which means that the gap thickness between "drift cathode-1st, GEM foil-2nd, GEM foil- 3rd, GEM foil- readout anode" with the unit "mm". The drift cathode and readout anode are made of PCBs. The main frame and inner frame are made of epoxy. The GEM foils have an effective area of 30 $cm \times 30$ $cm$, the upper copper layer is divided into 10 segments with each segment connected to a HV line via a 10 MΩ protection resistor. This design avoids possible damage to GEM foils from discharging and ensure any shorted segments are isolated from the rest, hence maximizing the operation efficiency of the detector.

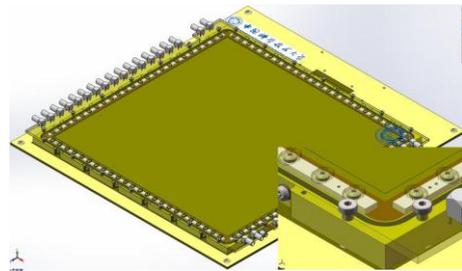

Fig. 3. Design of 30 $cm \times 30$ $cm$ GEM detector prototype, it has a "3-2-2-2" structure, the drift cathode and readout anode are made of PCBs. There are blind vias on the readout PCB which are used for providing space to the screws on inner frames.

There are 4 steps for the detector assembling: 1) assemble the three GEM foils on the inner frames. 2) put the main frame onto the drift PCB and fix it. 3) stretch the framed GEM foils by tightening the screws. 4) put the readout PCB on the main frame and fix it.

Submitted to Chinese Physics C

The HV divider is an 8-pin integrated resistor network with the 1st pin connected to the drift cathode, the 2nd to 7th pins connected to the 6 electrodes of the 3 GEM foils, and the last pin connected to the ground. The power dissipation in the HV divider due to the rather large current passing though the divider could heat up the metal pins and pose a potential danger of having the pins unsoldered. To assess this dangerous effect, a thermal sensitive camera was deployed to monitor the temperature of the pins while high voltage was applied to the divider. The results showed that the temperature of the pins kept below 100°C even when the input high voltage to the divider reached 4500 V which was higher than the nominal working voltage of the GEM detector by around 400 Volts. The pins are thus deemed to be safe in the presence of the power dissipation of the HV divider.

The high voltage from a power supply is delivered to the HV divider through a RC-low-pass filter composed of 100k$\Omega$ resistors and 2.2 nF capacitors to bypass high frequency noise coupled to the high voltage and hence improve the signal to background ratio of the GEM detector.

The readout PCB was designed specially for gain measurement. To make sure the charge produced in each event can be completely collected, the active area of the readout PCB is divided in 36 square sectors with the size of 5 $cm \times 5\ cm$ each. And each sector is read out independently.

## 4 Test setup and results

The GEM detector was mounted on an aluminum support plate. The whole detector was then placed in a copper-shielding chamber for tests with X-rays. There are 36 blind holes (radius=25 mm) on the outer side of the drift PCB. They were used as the window for the 8 keV X-rays produced by the Oxford Cu-target X-ray generator. The use of these holes significantly reduced the impact caused by the interactions of the X rays and PCB material. The X-ray generator was fixed on a platform which was able to move flexibly. The detector operated with the gas mixture of $Ar/CO_2$ (70/30) and high voltage of 4000 V. The signals from the detector were amplified by a charge sensitive pre-amplifier with a sensitivity of 0.8 V/pC and then shaped by an Ortec 671 amplifier. The output from the Ortec 671 was fed to a Multi-Channel Analyzer (Amptek MCA8000D) for energy spectrum measurement.

Fig.4(1) shows a typical signal from the pre-amplifier. The signal has an amplitude of about 30 mV with an exPonential tail spanning 5$\mu$ s. Fig.4(2) shows a typical energy spectrum measured by the GEM detector system of the Cu 8 keV characteristic X-ray produced by the X-ray generator. Both the full energy peak and the Argon escape-peak can be seen clearly on the spectrum. The energy resolution is determined to be about 20% using the full energy peak.

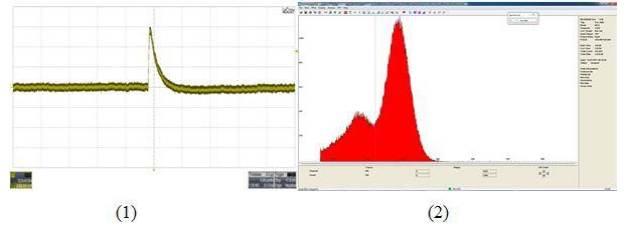

Fig. 4. (1) A typical signal from the charge sensitive pre-amplifier, it has an amplitude of 30 mV and an exponential tail spanning 5 $\mu$ s. (2) A typical energy spectrum of the Cu 8 keV characteristic X-ray, both full energy peak and Argon escape-peak can be seen clearly, the energy resolution is about 20%.

If both the average current of the signals produced by the GEM detector and the rate of the incident X-rays are known, the effective gas gain of the GEM detector can be determined by the following equation:

$$G = \frac{I}{R \times e \times 8keV/26.4eV}. \quad (1)$$

Where: I is the average current of the GEM output signal; R is the rate of X-ray; e is electron charge(1.6 $\times$ 10$^{-19}$ C), 8 keV is the energy of X-ray and 26.4 eV is the average ionization energy of the gas mixture of $Ar/CO_2$ (70/30), respectively.

In the X-ray test, the current collected from the readout PCB was directly measured by a Keithley6487 Picoammeter. The X-ray rate was estimated by counting the signals from the shaping amplifier with a CAEN N1145 scaler. Fig.5 shows the effective gas gain of the detector as a function of the high voltage applied to the detector. A clear exponential relation between the gain and the high voltage can be seen. The maximum gain can reach $10^5$ when the applied high voltage is 4.4 kV. The gain test was performed for the 35 individual GEM sectors. The 31st sector behaved in an unstable manner with a high voltage applied, and was finally excluded from the gain test. The unstable behavior was found to have been caused by improper cleaning operation on the sector during the detector assembling process.

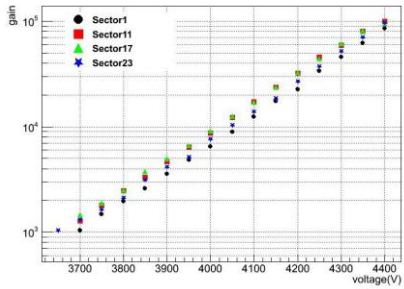

Fig. 5. Gain .VS. HV for Sector1、11、17 and 23

The gain of the GEM detector are shown in fig.6(1). The variation of effective gas gain is less than 31% with a working HV of 4 kV. And the variation reduces down to 21% when excluding sector 25 at the edge of the detector. Considering the edge effect, if the sector25 is excluded, the variation will be less than 21%. Fig.6(2) shows the uniformity of energy resolution with a HV of 4 kV. The energy resolution over the whole active area varies from 18% to 26%.

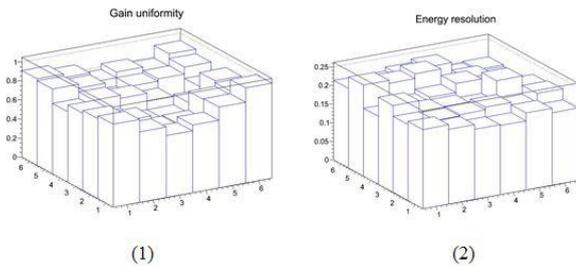

Fig. 6. (1) Uniformity of the effective gas gain at 4 kV, the gain variation is less than 31% and will be reduced down to 21% if sector 25 is excluded. (2) Uniformity of energy resolution, it varies from 18% to 26%.

## 5 Conclusions & Outlook

The NS2 self-stretch technique offers us a fast GEM assembling procedure producing a flexible GEM structure. The advantages of NS2 are summarized below:

1. The assembling procedure is very easy and fast, a 30 $cm \times 30$ $cm$ detector can be assembled in 1 hour；

2. All the GEM foils are self-stretched with no spacers needed producing no dead area inside the active area;

3. Any detector parts are replaceable. So malfunctioning GEM detectors can be repaired quickly with low cost.

A 30 $cm \times 30$ $cm$ GEM detector built with the NS2 technique was tested for effective gas gain and energy resolution. Very good uniformity of effective gain was observed over the whole active area. As the first large-area GEM detector, this GEM prototype serves as a valuable reference for the design of even larger GEM detector prototypes. In the future, the active area of GEM detector built with NS2 technique can even reach 1 $m \times 0.6$ $m$. These large-area GEM detectors will be a good candidate for tracking and triggering in high energy physics experiment, computer tomography in medical imaging, radiation monitoring in nuclear experiment, etc.

## 6 Acknowledgement

We would like to thank the CMS GEM collaboration and CERN PCB factory for supplying the GEM foils and allowing us to join their full-scale GEM R&D work. Special thanks to Archana Sharma, Rui De Oliveira and Leszek Ropelewsk for helpful discussions and suggestions.